# Time-evolving Impact of Trees on Street Canyon Microclimate

**Haiwei Li**[1,2], **Yongling Zhao**[1,4], **Ronita Bardhan**[2], **Aytac Kubilay**[1], **Dominique Derome**[3], **Jan Carmeliet**[1]

[1] Department of Mechanical and Process Engineering, ETH Zürich, Zürich, Switzerland

[2] Department of Architecture, University of Cambridge, Cambridge, UK

[3] Department of Civil and Building Engineering, Université de Sherbrooke, Sherbrooke, Canada

[4] Email: yozhao@ethz.ch

**Abstract.** Nowadays, cities are frequently exposed to heatwaves, worsening the outdoor thermal comfort and increasing cooling energy demand in summer. Urban forestry is seen as one of the viable and preferable solutions to combating extreme heat events and urban heat island (UHI) in times of climate change. While many cities have initiated tree-planting programmes in recent years, the evolving impact of trees on street microclimate, in a time span of up to several decades, remains unclear. We investigate the cooling effects of linden trees in five groups, i.e., 10-20, 20-30, 30-40, 40-60, and 60-100 years old. The leaf area index (LAI) and leaf area density (LAD) vary nonlinearly as the trees grow, peaking at different ages. Computational fluid dynamics (CFD) simulations solving microclimate are performed for an idealized street canyon with trees of varied age groups. Turbulent airflow, heat and moisture transport, shortwave and longwave radiation, shading and transpiration are fully coupled and solved in OpenFOAM. The meteorological data, including air temperature, wind speed, moisture, and shortwave radiation of the heatwave in Zurich (June 2019), are applied as boundary conditions. The results show that young trees in the age group of 10-20 years old provide little heat mitigation at the pedestrian level in an extreme heat event. Optimal heat mitigation by trees is observed for the group of 30-60 years old trees. Finally, the potential impact of growing trees as a heat mitigation measure on air ventilation is evaluated.

## 1. Introduction
The percentage of inhabitants living in cities is expected to rise from about 50% in 2010 to nearly 70% in 2050 [1]. Increasing urbanization, population densification and climate change raise many problems for urban microclimate. Air temperature in urban areas is usually higher than the air temperature in surrounding rural areas, which is defined as the urban heat island (UHI) effect [2]. The rise in urban temperature increases cooling energy demand, the concentration of airborne pollutants, and heat-related illness and mortality. Urban forestry is recognized as one of the vital sustainable measures for UHI mitigation [3], as it has enormous environmental benefits, such as reducing air and surface temperature [4], reducing air pollution and flood damage [5,6], buffering traffic noises [7], and increasing urban biodiversity [8]. Furthermore, the implementation of greenery also adds aesthetic advantages to cities and improves mental and physical health of social communities.

Vegetation alters the thermal and wind condition of the urban microclimate through several coupled multi-physical mechanisms, including shading, radiation trapping, evapotranspiration, and aerodynamic influence. As trees grow with time, these mechanisms vary and play different roles as a consequence of the changing vegetation properties, for instance, tree size, height, leaf area density (LAD), and leaf area index (LAI).

In terms of radiation and transpiration, the effects of trees often follow a diurnal cycle. During the daytime, the foliage of street trees provides shading and absorbs radiation but decreases the sky view factor of the canyon. At night, the trees may trap the longwave radiation under the tree canopy. LAD, LAI, and the dimensions of the tree foliage, especially the crown width, are significant factors that may influence the tree shades and the surrounding heat balance. A study using thermal satellite imagery in Terre Haute, Indiana, USA, showed that for every unit increase of LAI, the surface temperature is reduced by 1.2 °C [9]. Another study in the city of Dresden, Germany, showed that the surface temperature is reduced as the leaf area density (LAD) increases [10]. However, the study of Hien and Jusuf [11] shows that mature and bulk trees may lead to nighttime warming as the longwave radiation trapping is increased. High LAI and LAD values can also lead to the high transpirative cooling potential of trees [12]. A field study in Munich, Germany, observed three times higher transpiration in Tilla cordata trees compared to the Robinia pseudoacacia [13]. The LAI of the former species is 30% higher than the latter one. In terms of the aerodynamic influences, the shape, porosity, and drag of the trees determine the aerodynamic properties and the airflow around the trees, altering the flow structures and the turbulence mixing. The structure of the foliage increases the aerodynamic resistance and forms a thick stagnant layer around the leaves [14]. Plants with high foliage density are seen to reduce the air ventilation in the street canyons and, therefore, affect the pollutant dispersion below the urban canopy [15,16].

Understanding the effects of tree growth is essential for optimizing their environmental benefits. However, few studies investigate the time-evolving impacts trees' properties during decade-long growth, such as the height, absolute and relative size, LAI and LAD of the foliage. This study models the transpirative cooling, shading and aerodynamic influences of two rows of linden trees in a street canyon during three days of 2019 heatwave in Zurich. Six scenarios are simulated using in-house code urbanMicroclimateFoam in OpenFOAM, where the growth of trees' crown dimensions, height and the change of LAI and LAD are characterized by a series of tree age ranges, i.e., 0, 10-20, 20-30, 30-40, 40-60, and 60-100 years old.

## 2. Methodology

*2.1. Numerical method*
The study is carried out with a numerical urban microclimate computational fluid dynamics (CFD) simulations model to obtain high-resolution aerodynamic and thermal data of the local microclimate in OpenFOAM. The airflow is solved in the air subdomain using Reynolds-averaged Navier-Stokes (RANS) with k-ε turbulence models, with the heat and moisture transport in air, taking into account buoyancy. The model also takes into account the shortwave and longwave radiative exchanges in a radiosity model based on the view factor approach [17]. In the solid subdomains that model urban materials, such as buildings and streets, the coupled heat and moisture transport (HAM) equations are solved. The solid subdomains are coupled with the air subdomain at the boundaries, enabling the modeling of the temperature and moisture storage and transport at the surfaces [18].

Trees are modeled as porous medium in air subdomain, using sink and source terms for momentum, moisture, temperature, and turbulence quantities. The LAD value of trees defines the aerodynamic influence. The buoyancy is calculated with the Boussinesq approximation. The leaf energy balance is solved by discretizing the foliage of trees into small volumes, where the radiative, latent and sensible heat fluxes are calculated [12]. It is assumed that the stomata always have enough water uptake from the moist soil. The stationary energy balance of a single leaf, with the heat fluxes at the leaf surface, are defined by equations (1-3):

$$q_{rad,l} - q_{lat,l} - q_{sen,l} = 0 \tag{1}$$
$$q_{lat,l} = L_v \, h_{c,m} \, (p_{v,l} - p_v) \tag{2}$$
$$q_{sen,l} = h_{c,h} \, (T_l - T) \tag{3}$$

Where $q_{rad,l}$ (W/m²) represents the radiative flux, $q_{lat,l}$ (W/m²) represents the latent heat flux and $q_{sen,l}$ (W/m²) represents the sensible heat flux. $L_v$ (2.5 × 10⁶ J/kg) is the constant latent heat of vaporization. $h_{c,m}$ (s/m) is the convective mass transfer coefficient (CMTC), $p_{v,l}$ (Pa) is vapor pressure inside the leaf, $p_v$ (Pa) is the ambient vapor pressure. $p_{v,l}$ is assumed to be the saturated vapor pressure at leaf temperature. $h_{c,h}$ (W/m²K) is the convective heat transfer coefficient (CHTC) at the leaf surface, $T_l$ (K) is the leaf surface temperature, $T$ (K) is the air temperature. Finally, the equation of leaf energy balance can be combined and solved iteratively with the $T_l$ by equation (4):

$$T_l = T + \frac{q_{rad,l} - q_{lat,l}}{h_{c,h}} \tag{4}$$

*2.2. Description of the case study*

The case setup in the computational domain has a dimension of 480 m (streamwise, $x$) × 480 m (spanwise, $y$) × 122 m (height, $z$). A single stand-alone street canyon (aspect ratio = 1) with two identical buildings, a street, and two rows of trees is modeled, as shown in figure 1. A computational grid is generated following a sensitivity analysis in order to select optimal cell refinement for a high-resolution and accurate simulation. The cells of the air subdomain are refined from 4 m at a distance of 60 m away from the buildings to under 0.4 m within a distance of 10 m from the buildings. The cells are further refined near and in the vegetation. Prism layers, smaller than 0.2 m, are added to the building and street surfaces. The total cell number in the air domain is approximately 1.9 million.

The boundary conditions follow the actual metrological data in the city of Zurich, Switzerland, during the heatwave on 25-27 June 2019. The hourly ambient temperature, humidity ratio, solar radiation parameters, and wind velocity magnitude data are used in the simulation. The incoming wind is in $x$ direction, and the wind direction is assumed to remain the same during the three days of simulation for simplicity. The profile of wind speed at the inlet follows a fully-developed atmospheric boundary layer, with the assumption of a neutral stratification, as defined in equations (5-7):

$$U(z) = \frac{u^*_{ABL}}{\kappa} \ln\left(\frac{z + z_0}{z_0}\right) \tag{5}$$

$$k(z) = \frac{{u^*_{ABL}}^2}{C_\mu^{0.5}} \tag{6}$$

$$\epsilon(z) = \frac{{u^*_{ABL}}^3}{\kappa(z + z_0)} \tag{7}$$

where $U(z)$ represents the horizontal wind speed at height z, $u^*_{ABL}$ represents the atmospheric boundary layer friction velocity, $z_0$ represents the aerodynamic roughness length, which is chosen as 1 m, $\kappa$ represents the von Karman constant and $C_\mu$ is a model constant, which equals to 0.09.

The vegetation properties change as the trees grow up. A common species of trees in Europe, small-leaved linden, is modeled, which is also named Tilla cordata Mill. As the linden trees grow up, the height of the trees can reach 30-35 m. LAD and LAI are not seen to have a linear growth with trees' age [19]. The height, crown diameter, LAD and LAI of the trees, among other properties, are provided in table 1 following the literature [20–23].

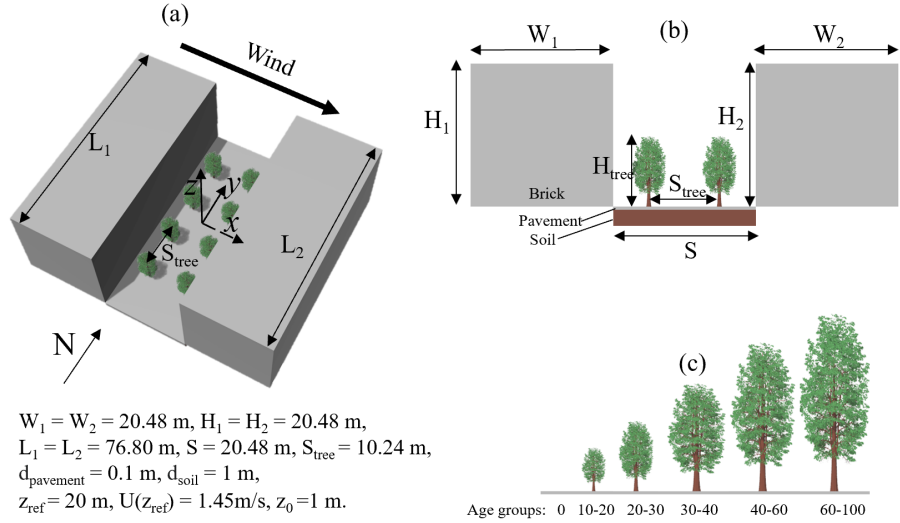

**Figure 1.** Case study setup in the CFD simulation. (a) and (b) demonstrate the implementation of street trees in the canyon. The $y$-axis direction is towards the north. The spacing of the trees ($S_{tree}$) is 10.24 m. The thickness of the pavement ($d_{pavement}$) and the thickness of the soil ($d_{soil}$) are 0.1 m and 1 m under the trees. The surface roughness length ($z_0$) is 1 m, which corresponds to a uniformly-built town. The horizontal wind velocity at approximate roof height ($z_{ref}$) is 1.45 m/s. (c) demonstrates the sizes of trees in 6 different scenarios.

**Table 1.** Detailed tree vegetation properties in 6 scenarios of age groups. $H_{tree}/H_1$ denotes the relative height of the trees to the height of the canyon, $H_{tree}$ denotes the height of the trees, $D_{crown}$ denotes the diameter of the crown, LAD is the leaf area density, LAI is the leaf area index, $a_{tree}$ is the albedo of the leaves, $r_{s,min}$ is the minimal stomata resistance of the leaves.

|   | Age (years) | $H_{tree}/H_1$ | $H_{tree}$ (m) | $D_{crown}$ (m) | LAD (m²/m³) | LAI (m²/m²) | $a_{tree}$ | $r_{s,min}$ (s/m) |
|---|---|---|---|---|---|---|---|---|
| 1 | 0 | 0 | 0 | 0 | 0 | 0 | 0.15 | 150 |
| 2 | 10-20 | 0.25 | 5.12 | 2.3 | 0.7 | 1.68 | 0.15 | 150 |
| 3 | 20-30 | 0.5 | 10.24 | 4.7 | 1.4 | 6.91 | 0.15 | 150 |
| 4 | 30-40 | 0.75 | 15.36 | 7 | 1.6 | 11.65 | 0.15 | 150 |
| 5 | 40-60 | 1 | 20.48 | 8.5 | 1.4 | 13.48 | 0.15 | 150 |
| 6 | 50-100 | 1.25 | 25.6 | 10 | 1.2 | 12.91 | 0.15 | 150 |

## 3. Results and discussion

The spatial-averaged air temperature and humidity ratio at pedestrian level (1.8 m height) in the street canyon are shown in figure 2. The scenario without trees has the highest air temperature at all times. The peak temperature is observed at 15:00 on the second day of the heatwave. With the implementation of street trees, the air temperature is overall largely reduced, especially during the daytime, when the shading and transpirative cooling collectively provide the cooling effects to the canyon. The highest air temperature reduction reaches 4 °C at the location under the trees, and 1.1 °C for the pedestrian level spatial-averaged reduction, at the peak temperature when the highest trees (60-100 years old) are present. For cases with trees, the spatial-averaged humidity ratio is also higher than the case without trees during the daytime due to the transpiration of leaves.

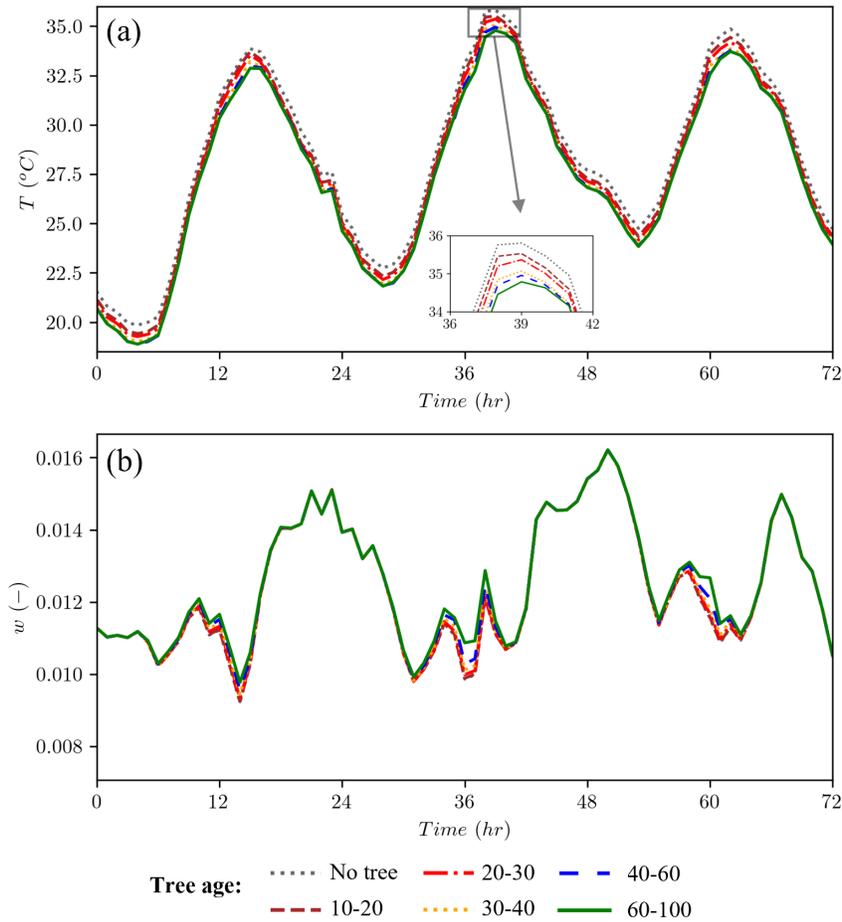

**Figure 2.** Pedestrian level spatial-averaged (a) air temperature ($T$, °C) and (b) humidity ratio ($w$, kg/kg) during three days of the heatwave. The peak temperature is observed at 15:00 on the second day of the heatwave. Six different lines represent six scenarios of tree age, from 0 to 100 years old.

Figure 3 presents the spatial profiles of the Universal Thermal Climate Index (UTCI, °C) on two vertical spanwise planes ($y - z$ planes) in the canyon at the peak temperature time, 15:00 on the second day of the heatwave, where figure 3($a_1$-$f_1$) are obtained in the center of the trees near the windward walls (plane 1) and figure 3($a_2$-$f_2$) are in the center of the canyon (plane 2). UTCI is an assessment of the thermophysiological effects on the atmospheric environment, which is calculated based on the metabolic rate, clothing insulation, air temperature, mean radiant temperature, air speed, and relative humidity [24]. The mean radiant temperature takes into account the shortwave and longwave radiation received by the human body.

The results show that trees of different ages bring a distinctly different level of cooling and influence on UTCI around trees. In plane 1, at the location of the above trees, the UTCI in the youngest trees scenario (figure 3$a_2$) is slightly higher than the UTCI in the no tree scenario (figure 3$a_1$). The UTCI reduction is also seen around the trees as the trees grow up. For plane 1, UTCI is overall reduced at the locations around and under the trees. On plane 2(figure 3$a_2$-$f_2$), the UTCI reduction at the pedestrian level is limited for tree age groups 10-20 and 20-30 (figure 3$b_2$-$c_2$), and the reduction becomes visible as the trees are older than 30 years old (figure 3$d_2$-$f_2$).

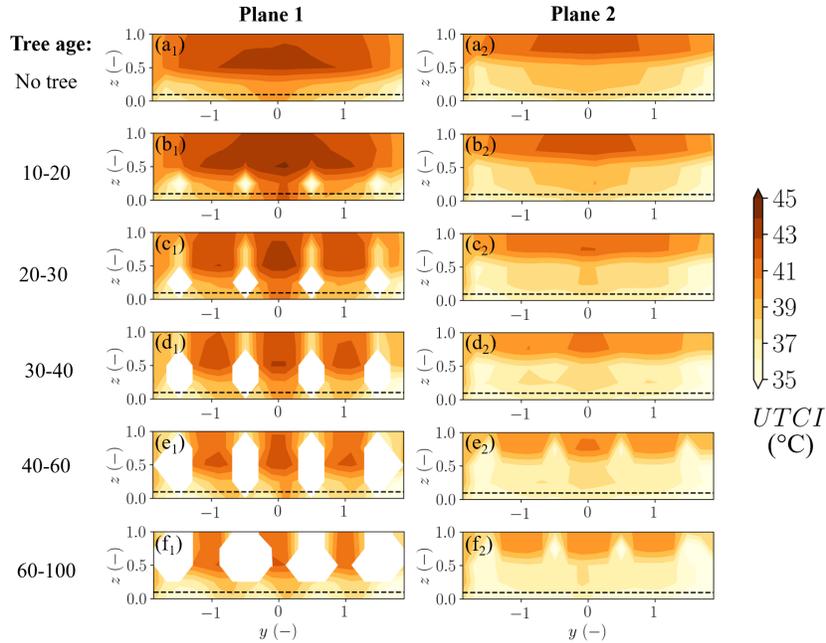

**Figure 3.** The Universal Thermal Climate Index (UTCI) contour distribution on vertical spanwise ($y - z$ plane), ($a_1$-$f_1$) plane 1, center of the trees near the windward wall, and ($a_2$-$f_2$) plane 2, the center of the canyon, at peak temperature time, 15:00 on the second day of the heatwave. The dashed line represents the pedestrian level height at 1.8 m.

The presence of street trees may also influences the air ventilation of the street canyon. Figure 4 shows the air ventilation rate at peak temperature hour, 15:00, on the second day of the heatwave for all the age group scenarios. The analysis of the ventilation rate at the canyon openings was used in the literature [14,25], where the air ventilation rate consists of air removal and air entrainment. It is seen in figure 4 that the air ventilation of the street canyon gets smaller as the trees grow. The larger foliage of trees causes a larger aerodynamic resistance in the canyon. The reduction in air removal is particularly higher than the reduction of air entrainment. When the largest trees are present, the ventilation can be reduced to less than half of the case without trees.

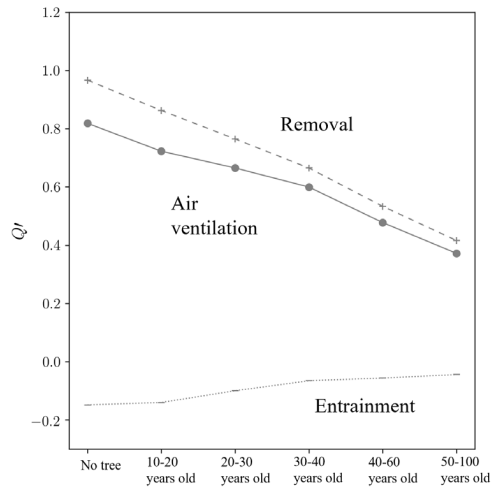

**Figure 4.** Normalized air ventilation rate from all openings (air removal and air entrainment) at peak temperature time, 15:00 on the second day of the heatwave. The normalized air ventilation

rate is calculated by the air ventilation rate normalized by $H_1 \times H_2 \times U(z_{ref})$, following literature [14,25]. Both air removal and air entrainment are reduced as the trees grow up.

## 4. Conclusion

Our study investigates a street canyon microclimate with trees in a range of age groups, using an urban microclimate CFD simulation model. The simulation results reveal that the tree properties can significantly affect their cooling performance during heatwaves. UTCI analysis and air ventilation rate analysis are performed to study the thermal comfort and air removal and entrainment in the canyon. Our results show that:

- As the trees grow taller and larger, the LAI and LAD are changed, and the cooling effects of the trees become more pronounced.
- Overall, the maximum air temperature reduction can reach 4 °C directly under the trees and 1.1 °C for the whole pedestrian-level spatial-averaged temperature reduction.
- At the pedestrian level, trees under 30 years old provide overall very limited cooling potential, while the shading and transpirative cooling effects of trees only become beneficial after reaching 30 years old.
- The presence of the largest trees, e.g., aged 60-100, significantly reduces the air ventilation of the canyon, which may cause adverse effects on pollutant dispersion and heat removal from the canyon by the wind.
- Urban planners should carefully select and manage street trees, in terms of their size, height, LAI, and LAD, to benefit the most from the trees. Proper simulations may carry out when planning the implementation of trees.

The simulation is simplified to center on the time-evolving impact of trees. A single street canyon morphology with an aspect ratio of 1 is studied. Moreover, the boundary condition of wind is simplified to flow always from west to east. In future work, we will investigate the time-evolving impact of trees in a realistic neighborhood model. The impacts of different tree properties will be studied systematically.